\documentstyle[preprint,aps]{revtex}
\title{The phase-dependent linear conductance of a superconducting
quantum  point contact}
\author{A. Levy Yeyati, A. Mart\'{\i}n-Rodero and J. C. Cuevas}
\address{
Departamento de F\'\i sica de la Materia Condensada C-XII.\\
Facultad de Ciencias. Universidad Aut\'onoma de Madrid.\\
E-28049 Madrid. Spain.}
\begin{document}

\draft
\maketitle

\begin{abstract}
The exact expression for the phase-dependent linear conductance of a
weakly damped superconducting quantum point contact is obtained. The
calculation is performed by summing up the complete perturbative series
in the coupling between the electrodes. The failure of any finite order
perturbative expansion in the limit of small voltage and small
quasi-particle damping is analyzed in detail.
In the low transmission regime this nonperturbative calculation
yields a result which is at variance with standard tunnel theory. Our
result predicts the correct sign of the quasi-particle pair interference term
and exhibits an unusual phase-dependence at low temperatures in
qualitative agreement with the available experimental data.
\end{abstract}

\vspace{1cm}

\narrowtext

Since the early stages in the theory of Josephson junctions it has
been customary to write the total current through the junction as

\begin{equation}
I = I_{J} \sin \phi + G_{0}(1+ \epsilon \cos \phi) V  ,
\end{equation}

\noindent
where the total superconducting phase difference between the electrodes
is related to the applied bias voltage by
$ d \phi/d \tau = 2 e V/ \hbar  = \omega_{0}$.

Eq. (1) was first derived from a microscopic model by Josephson
\cite{Joseph}. An equivalent expression has been widely used to describe
superconducting point contacts \cite{Likharev}.
The second term in Eq. (1) defines a phase-dependent
conductance, $ G(\phi) = G_{0}(1+ \epsilon \cos \phi)$, whose existence
was confirmed by a series of experiments during the seventies
\cite{exp}. However, it soon became apparent that the experimental
results for low temperatures seemed to be best fitted by Eq. (1) with
a value of the parameter $\epsilon \approx -1$, while tunnel theory
predicts $\epsilon =+1$ in the limit $T,V \rightarrow 0$
\cite{Barone}.
Furthermore, an experiment measuring the complete phase dependence
of the linear conductance in a point contact
showed a strong departure of $G(\phi)$ from
the simple $\cos \phi$-like form \cite{Rifkin}, given by tunnel theory.
There was at the time a large number of theoretical
works trying to explain the discrepancy between tunnel theory and
experiments, most of them relying on the introduction of a
phenomenological broadening of the Riedel peak (for a review see
ref. \cite{Barone}). However, a completely satisfactory explanation of
this issue seems still to be lacking.

  On the other hand, there has been in the last few years a renewed
interest on the theory of superconducting weak links associated with
the increasing technological capability for the fabrication of
nano-scale superconducting devices. This opens the possibility of a
closer comparison between theoretical predictions and clean experiments
even on a nearly atomic scale \cite{van}. It therefore seems
appropriate  at this point to perform
a careful re-examination of the transport properties in
superconducting point contacts going beyond the limits of
standard tunnel theory.
In this direction, one should mention some important theoretical
contributions
\cite{OBTK,Arnold}, clarifying the crucial role played by multiple
Andreev-reflections in the explanation of the sub-harmonic structure
in superconducting point contacts. This sub-gap structure
can be currently analyzed experimentally with increasing resolution
\cite{van,Klein}.

As will be shown in this letter, the presence of edge singularities
in the spectral densities of the superconductors leads to an enhanced
weight of these multiple scattering processes. As a consequence, a
nonperturbative calculation is needed to obtain the correct result for the
phase-dependent conductance. In a non-biased junction,
the summation of the infinite series of multiple scattering events leads to
the appearance of bound states inside the superconducting gap, whose
existence has been discussed
in previous theoretical works \cite{Kulik,Beenakker,us}.
In the same way, the presence of bound states will allow us to obtain an
exact expression for $G(\phi)$ valid in the limit of small
quasiparticle damping.

We consider for simplicity the case of a symmetrical contact. Then,
with $L$ and $R$ representing the left and right electrodes, we have
$|\Delta_{L}| = |\Delta_{R}| = \Delta$ and $\phi = \phi_{L} - \phi_{R}$,
where $\Delta$ is the modulus of the superconducting order parameter,
$\phi$ representing the total phase difference which is supposed to
drop abruptly at the interface between both electrodes.

To describe the biased point contact we use the following model
Hamiltonian

\begin{equation}
\hat{H} = \hat{H}_{L} + \hat{H}_{R}
+ \hat{H}_{LR} + \frac{eV}{2} \left(\hat{N}_L - \hat{N}_R \right) ,
\end{equation}

\noindent
where $\hat{H}_{L}$ and $\hat{H}_{R}$ are the BCS Hamiltonians for the
uncoupled electrodes,
($\hat{N}_{L}, \hat{N}_{R}$) being the corresponding
total number operators, and $eV$ is the applied
bias voltage. The term $\hat{H}_{LR}$ coupling both electrodes is
assumed to have the following form

\begin{equation}
\hat{H}_{LR} = \sum_{i,j,\sigma} ( t_{i j}
c^{\dagger}_{i \sigma} c_{j \sigma}
+ t_{j i} c^{\dagger}_{j \sigma} c_{i \sigma})  ,
\end{equation}

\noindent
where ($i,j$) stand for orbitals on the
(left, right) electrodes respectively. For the present analysis it is
convenient to consider first the simplest case in which there is a
single channel connecting both electrodes (the multi-channel case will
be briefly discussed at the end of the letter). In this case ($i
\equiv L, j \equiv R$) denote the two orbitals connected by the single
hopping element $t_{LR}=t$. We perform the standard unitary
transformation \cite{Ryck,Scalap},
by means of which the  system dynamics is governed by
the following time-dependent Hamiltonian

\begin{equation}
\hat{H}(\tau) = \hat{H}_{L} + \hat{H}_{R} +\sum_{\sigma} ( t
e^{i \phi(\tau)/2} c^{\dagger}_{L \sigma} c_{R \sigma} + t
e^{-i \phi(\tau)/2} c^{\dagger}_{R \sigma} c_{L \sigma}) ,
\end{equation}

\noindent
where $\phi(\tau) = \phi_{0} + 2eV \tau/\hbar $. Within this representation
the time-dependent phase only appears in the phase factors multiplying the
hopping elements. The transport properties of this system can be analyzed
using nonequilibrium Green functions techniques \cite{Arnold,us,Zaitsev}
with the time-dependent coupling term treated as a perturbation. The
most relevant quantity in this formalism is the nonequilibrium
distribution function $G^{+,-}$, which in a superconducting broken
symmetry (Nambu) representation is defined by

\begin{equation}
\hat{G}^{+-}_{i,j}(\tau,\tau^{\prime})= i \left(
\begin{array}{cc}
<c^{\dagger}_{j \uparrow}
(\tau^{\prime}) c_{i \uparrow}(\tau)>   &
<c_{j \downarrow}(\tau^{\prime}) c_{i \uparrow}(\tau)>  \\
<c^{\dagger}_{j \uparrow}(\tau^{\prime})
c^{\dagger}_{i \downarrow}(\tau)>  &
<c_{j \downarrow}(\tau^{\prime})
 c^{\dagger}_{i \downarrow}(\tau)>
\end{array}  \right) .
\end{equation}

In terms of these functions the current through the contact can be
written as

\begin{equation}
I(\tau) =\frac{2 e}{\hbar} \left[\hat{t}(\tau) \hat{G}^{+-}_{RL}(\tau,\tau) -
\hat{t}^\dagger(\tau) \hat{G}^{+-}_{LR}(\tau,\tau) \right]_{11}     ,
\end{equation}

\noindent
where $\hat{t}$ is the matrix hopping element in the Nambu
representation

\begin{equation}
\hat{t} = \left(
\begin{array}{cc}
 t e^{i \phi(\tau)/2}  &   0      \\
  0                    &   -t e^{-i \phi(\tau)/2}
\end{array} \right)  .
\end{equation}

Within this perturbative approach the tunnel theory expression for
the current (Eq. (1)), can be obtained at the lowest-order
in $\hat{H}_{LR}$.
The conductance $G_0 ( 1 + \epsilon \cos \phi)$ thus obtained becomes
a divergent quantity in the limit $V \rightarrow 0$ \cite{Barone}.
In order to ensure the existence of a linear regime, a finite energy
relaxation rate $\eta$ must be introduced into this superconducting
mean field theory ($\eta$ represents
the damping of the quasi-particle states, which in a real system is
always present due to inelastic scattering processes).
As we shall see, according to the value of $\eta$ and the
normal transmission coefficient of the junction, $\alpha$
\cite{comment1},
two different regimes can be identified: the weakly damped regime,
for which $\eta \ll \alpha \Delta$ and the strongly damped case,
where $\eta \gg \alpha \Delta$. In this work we are mostly concerned
with the analysis of the first regime, where the most interesting
effects appear.

A remarkable fact about the perturbative expansion
in the weakly damped situation is that
contributions corresponding to higher order processes turn out to be
increasingly divergent in the zero bias limit \cite{Hasselberg}.
In particular, it can
be easily demonstrated that contributions to the total current of
order $t^{2n}$, $n \geq 2$, diverge like $ \sim t^{2n}/ \eta^{n-1}$
(the lowest order contribution diverges as $\sim t^2 \ln \eta$).
This result is a direct consequence of the increasing contribution from
the superconducting gap edges singularities. Therefore, a correct
answer cannot be found in principle by means of a finite order
perturbative expansion.

One could draw a formal analogy with the case of a high-density electron gas,
where the diagrammatic expansion in the bare Coulomb potential is
also increasingly divergent. As in that case, the solution can be found
by ``dressing" the perturbative potential, i.e. $\hat{H}_{LR}$.
In the present problem
the dressed quantities (left-right coupling, propagators)
can be exactly obtained in the zero voltage limit
 by evaluating the complete perturbative series.
To this end, we find it convenient to express all quantities in terms of a
renormalized left-right coupling element which satisfies the following
Dyson equation

\begin{equation}
\hat{T}^{a,r}(\tau,\tau^{\prime}) = \hat{t}(\tau) \delta(\tau-
\tau^{\prime})+ \hat{t}(\tau) \hat{g}^{a,r}_{R}(\tau-\tau_{1})
\hat{t}^{\dagger}(\tau_{1})
\hat{g}^{a,r}_{L}(\tau_{1}-\tau_{2}) \hat{T}^{a,r}(\tau_{2},
\tau^{\prime}) ,
\end{equation}
\noindent
where $\hat{g}^{a,r}_{L}$ and $\hat{g}^{a,r}_{R}$ represent the
(advanced, retarded) Green functions of the uncoupled left and right
electrodes respectively (integration over internal times is implicitly
assumed).
{}From Eq. (8), the relation between the
renormalized coupling $\hat{T}$ and the exact (advanced and retarded)
Green functions is easy to obtain.
In the same way, the nonequilibrium distribution function
$\hat{G}^{+-}$,
which is related to $\hat{G}^r$ and $\hat{G}^a$,
can be written in terms of $\hat{T}$ \cite{ALY}.

Integral equations like Eq. (8) adopt a simpler form
when Fourier transformed with respect to their temporal arguments
\cite{Arnold,ALY}.
Defining the Fourier components $\hat{T}_{n,m}(\omega)$ as

\begin{equation}
\hat{T}_{n,m}(\omega) = \int d\tau \int d\tau^{\prime}
e^{-i \left(n\phi(\tau) - m\phi(\tau^{\prime}) \right)/2}
e^{-i \omega (\tau-\tau^{\prime})}
\hat{T}(\tau,\tau^{\prime}) ,
\end{equation}

\noindent
the total current can then be expressed in the form
$ I(\tau) = \sum_m I_m \exp{im \phi(\tau)/2}$,
where the complex coefficients, $I_m$, do not depend on $\phi(\tau)$
and are given by

\begin{eqnarray}
I_m &= & \frac{2 e}{h} \int d\omega \sum_n
\left[ \hat{T}_{0,n}^r \hat{g}_R^{+-}(\omega + n \frac{\omega_0}{2})
\hat{T}_{n,m}^{r
\dagger} \hat{g}_L^a(\omega + m \frac{\omega_0}{2})
- \hat{g}_L^{r}(\omega) \hat{T}_{0,n}^r
\hat{g}_R^{+-}(\omega + n \frac{\omega_0}{2}) \hat{T}_{n,m}^{r \dagger}
\nonumber \right. \\
&& \left. + \hat{g}_R^{r}(\omega) \hat{T}_{0,n}^{a \dagger}
\hat{g}_L^{+-}(\omega + n \frac{\omega_0}{2}) \hat{T}_{n,m}^{a}
 - \hat{T}_{0,n}^{a \dagger}
 \hat{g}_L^{+-}(\omega + n \frac{\omega_0}{2})
\hat{T}_{n,m}^{a}
\hat{g}_R^{a}(\omega + m \frac{\omega_0}{2}) \right]_{11} .
\end{eqnarray}

It can be seen from Eq. (8) that $T_{n,m}=0$ for even $n-m$ and
therefore only even Fourier components of the current are different
from zero.

For the following analysis it is useful
to divide the total current into dissipative and
nondissipative contributions. The supercurrent part, given by $I_{S} =
-2  \sum_{m > 0}  \mbox{Im}(I_m)  \sin [m \phi(\tau)]$,
 tends to a finite value in the limit $V \rightarrow  0$.
On the other hand,
the dissipative part is given by $I_{D}  =  I_{0}  +  2  \sum_{m >0}
\mbox{Re}(I_{m}) \cos [m \phi(\tau)]$, and goes to zero as $I_{D}
\sim G(\phi) V$, $G(\phi)$ being the zero voltage conductance. The
linear term can be straightforwardly derived from Eq. (10) by expanding
the Fermi functions appearing in $\hat{g}_{L,R}^{+-}$ \cite{comment2}
up to first order in $V$ and evaluating the
remaining factors at zero voltage.

In this limit the Fourier components
satisfy $\hat{T}_{n,m}=\hat{T}_{0,m-n} \equiv \hat{T}_{m-n}$,
and can be shown to obey the simple recursive relations

\begin{eqnarray}
\hat{T}_{n+2}(\omega) & = & z(w) \hat{T}_{n}(\omega) \nonumber  \\
\hat{T}_{-n-2}(\omega) & = & z(\omega) \hat{T}_{-n}(\omega) \hspace{1cm}
(n \geq 1) ,
\end{eqnarray}

\noindent
where $z(\omega)$ is a scalar complex function. In the weakly
damped regime and within the
energy interval $\Delta > |\omega| > \Delta \sqrt{1-\alpha}$
this function reduces to a phase factor
$z(\omega) = \exp{i \varphi(\omega)}$, where

\begin{equation}
\varphi(\omega) = \arcsin \left(\frac{2}{\alpha \Delta^2}
\sqrt{\Delta^2-\omega^2}
\sqrt{\omega^2-(1-\alpha) \Delta^2} \right) .
\end{equation}

This clearly shows that in the weakly damped regime
and within this energy interval all
multiple scattering processes become equally important.
Therefore,
all Fourier components contribute to the renormalized coupling in this
region, giving rise to singularities which can be shown to be associated
with the existence of interface bound states. In fact, the renormalized
coupling in this energy region can be easily obtained from Eqs. (11)
and (12), giving

\begin{equation}
\sum_{n} \hat{T}_{n}(\omega) e^{in \phi/2} = \frac{\hat{T}_{1}(\omega)
 e^{i\phi/2}e^{i(\varphi+\phi)} }
 {1-e^{i(\varphi+\phi)} }  + \frac{\hat{T}_{-1}
 (\omega) e^{-i \phi/2}e^{i(\varphi-\phi)} }{1-e^{i(\varphi-\phi)} } ,
\end{equation}

\noindent
which exhibits singularities at $\varphi(\omega)= \pm \phi$. From Eq. (12)
it follows that these singularities correspond to simple poles at
$\omega_{S} = \pm \Delta \sqrt{1-\alpha \sin^2(\phi /2)}$. These
are the interface bound states inside the gap
of a superconducting point contact, as derived by different
authors \cite{Kulik,Beenakker,us}.

In the same way, the complete harmonic series must be
evaluated in order to obtain the contributions to both the dissipative
and nondissipative parts of the current coming from the energy range
$\Delta>|\omega| > \Delta \sqrt{1-\alpha}$.
Again, these infinite summations can be easily performed making use
of the recursive relations of Eq. (11).
It is then found that the integrand for both
parts of the current becomes singular at $\omega = \pm \omega_S$.
The contribution of these poles yields

\begin{equation}
I_{S}(\phi) =  \frac{e \Delta}{2 \hbar}
\frac{\alpha \sin \phi}{\sqrt{1-\alpha \sin^2(\phi/2)} }
\tanh (\frac{\beta \omega_{S}}{2} )
\end{equation}
\noindent
and
\begin{equation}
I_{D}(\phi) = \frac{2e^2}{h} \frac{\pi}{16 \eta} \left[ \frac{\Delta \alpha
 \sin \phi}{\sqrt{1-\alpha \sin^2 (\phi/2)} } \mbox{sech}
(\frac{\beta \omega_{S}} {2}) \right]^2 \beta V .
\end{equation}

In Eq. (14) the previously known result for the zero bias supercurrent
is recovered \cite{Kulik,Beenakker,us}. The expression for the dissipative
current given above is the main result of this letter. The linear
conductance thus obtained can be seen to depend on $\eta$ as $\sim
1/\eta$, i.e. proportional to a relaxation time. This result seems more
physically sound than that of tunnel theory which predicts
a dependence $\sim \ln \eta$.
Notice that in the low barrier transparency regime Eq. (15) depends on the
transmission coefficient as $\alpha^2$ which means that Andreev reflection
processes dominates over single quasiparticle tunneling in the
zero voltage limit.

It is also worth commenting that Eq. (15) can be derived
in a different way, by relating the linear conductance to the
equilibrium current fluctuations via the fluctuation-
dissipation theorem, giving further support for the validity
of this expression. This will be the subject of a forthcoming
publication.

Our theory yields a phase-dependent linear conductance which
strongly deviates from the tunnel theory result of Eq. (1).
In the limit of low barrier transparency, Eq. (15) predicts
$G(\phi) \sim 1 - \cos(2 \phi)$ instead of $G(\phi) \sim
1 + \epsilon \cos \phi$ of standard tunnel theory. Therefore, the
linear conductance in Eq. (1) can never be recovered in the weakly
damped regime. On the other hand, with increasing values of $\eta$
multiple scattering processes are progressively damped (the function
$z(\omega)$ is no longer a phase factor decaying exponentially with
$\eta$); eventually, when $\eta \gg \alpha \Delta$
only the lowest-order processes contribute to the current and
Eq. (1) is recovered. This explains the discrepancy between tunnel
theory and the experiments (usually referred to as the ``$\cos \phi$
problem''), because the experimental conditions should correspond to
the weakly damped case in order that the Josephson effects could be
observed \cite{comment3}.

Another interesting limiting case of Eq. (15) corresponds to the
ballistic, i.e. $\alpha \rightarrow 1$, regime. In this case and
for large temperatures
$G(\phi)$ behaves approximately as $(1 - \cos{\phi})$, in agreement with
the result given by Zaitsev \cite{Zaitsev}.
However, the most unusual phase-dependence of $G(\phi)$ appears
for high values of the transmission and low temperatures
($k_BT < \Delta$).
This is illustrated in Fig. 1,
where $G(\phi)$ is plotted for two different temperatures and increasing
values of the transmission. The only experiment where the full phase
dependence of $G(\phi)$ was measured is, to our knowledge, that of
ref. \cite{Rifkin}. Their measured $G(\phi)$ strongly deviates from a
$\cos \phi$-like form, being almost negligible for small values of
$\phi$ and exhibiting a large increase around $\phi \sim \pi/2$.
As can be observed in Fig. 1, this behavior is in qualitative
agreement with our results at any given temperature for sufficiently
large transmission. However, a detailed comparison should require a
more exhaustive experimental study of $G(\phi)$ for different barrier
transparencies and temperature regimes. We believe that these
measurements are now becoming feasible with recent advances in the
fabrication of nanoscale superconducting contacts.

The multi-channel generalization of our
results is formally straightforward.
For a general contact geometry it would lead to a superposition of
contributions like those of Eqs. (14)
and (15) for each transverse mode, which
can in principle have different transmission probabilities. We do not
expect that this effective
averaging process would alter in a significant way the
phase dependence of Eqs. (14) and (15).

In conclusion, it has been shown that a nonperturbative
calculation is needed for obtaining the total current through a
weakly damped superconducting point contact in the linear regime. Using
a simple model Hamiltonian we are able to obtain exactly
the phase-dependent linear conductance. The resulting expression
is in good agreement with the available experimental
data and we believe it can
provide a motivation for more detailed experimental studies.

\acknowledgements
Support by Spanish CICYT (Contract No. PB93-0260) is acknowledged.
One of us (A.L.Y.) acknowledges support by the European
Community under contract No.CI1*CT93-0247.
The authors are indebted to F.J. Garc\'{\i}a-Vidal,
F. Flores, N. Majlis and F. Sols for stimulating discussions.

\begin{figure}
\caption{Phase dependence of the linear conductance given by Eq. (15)
for two different temperatures and increasing values of the normal
transmission coefficient ($G(\phi)$ is normalized to its maximum value).}
\end{figure}


\begin{references}
\bibitem{comment2} Notice that the equilibrium distribution functions
for the uncoupled electrodes are simply given by $\hat{g}^{+-}_{L,R}
(\omega) = f(\omega)(\hat{g}^a_{L,R}(\omega) -
\hat{g}^r_{L,R}(\omega))$, where $f(\omega)$ is the Fermi function.
\bibitem{comment3} The actual observation of the Josephson effects is
usually hindered by thermal fluctuations and it is therefore necessary
to have large enough barrier transparencies or sufficiently low temperatures.
\end{references}
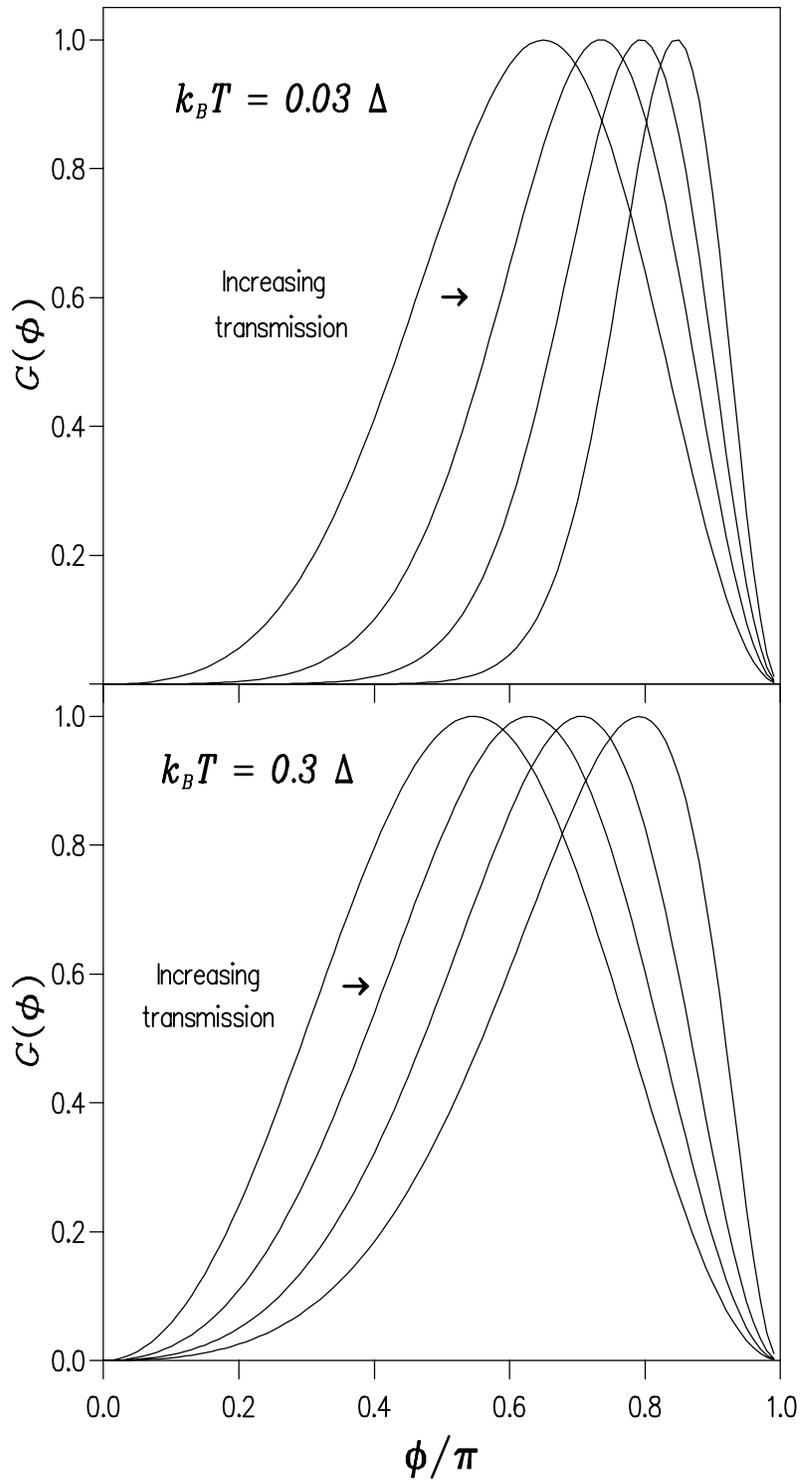
\end{document}